\documentclass{sig-alternate} 
\usepackage{tabularx}
\usepackage{booktabs}
\usepackage{datetime}
\usepackage{subfigure}
\usepackage{url}
\usepackage{paralist}
\usepackage{algorithm}
\usepackage[noend]{algpseudocode}
\usepackage{epstopdf}
\usepackage{flushend}
\usepackage{lipsum}
\usepackage{ntheorem}

\permission{Permission to make digital or hard copies of all or part of this work for personal or classroom use is granted without fee provided that copies are not made or distributed for profit or commercial advantage and that copies bear this notice and the full citation on the first page. Copyrights for components of this work owned by others than ACM must be honored. Abstracting with credit is permitted. To copy otherwise, or republish, to post on servers or to redistribute to lists, requires prior specific permission and/or a fee. Request permissions from Permissions@acm.org.}
\conferenceinfo{WebSci '15}{June 28 - July 01, 2015, Oxford, United Kingdom}
\CopyrightYear{2015}
\crdata{978-1-4503-3672-7/15/06\\
http://dx.doi.org/10.1145/2786451.2786460}

\begin{document}

\newcommand{\seq}[1]{\langle\,{#1}\,\rangle}
\newcommand{\set}[1]{\left\{\,{#1}\,\right\}}
\newcommand{\bigset}[1]{\left\{\,{#1}\,\right\}}
\newcommand{\bigtuple}[1]{\left(\,{#1}\,\right)}
\newcommand{\card}[1]{\left|\,{#1}\,\right|}
\newcommand{\tup}[1]{\left(\,{#1}\,\right)}
\newcommand{\argmin}[2]{\underset{#1}{\operatorname{arg\,min}}{\:\: #2}}
\newcommand{\stdmin}[2]{\underset{#1}{\operatorname{min}}{\:\: #2}}
\newcommand{\simplemin}[1]{\ensuremath{\underset{#1}{min}\;}} 
\newcommand{\pow}[2]{\ensuremath{#1^#2\;}}
\newcommand{\comment}[1]{}
\newcommand{\ra}[1]{\renewcommand{\arraystretch}{#1}}
\newcommand {\N}{\mathcal{N}}
\newcommand{\argmax}{\operatornamewithlimits{argmax}}

\hyphenation{Map-Reduce}
\hyphenation{opti-mi-za-tion}
\hyphenation{Wiki-pedia}
\hyphenation{ment-ioned}

\newtheorem*{entity_lag*}{Entity Lag:}
\newtheorem*{emerging_entities*}{Emerging Entities:}
\newtheorem*{ee_ep*}{Emerging Entities in Event Pages:}
\newtheorem*{news_density*}{News Reference Density:}
\newtheorem*{event*}{Event:}
\newtheorem*{entity*}{Entity:}

\title{How much is Wikipedia Lagging Behind News?}

\numberofauthors{1}
\author{
Besnik Fetahu, Abhijit Anand, Avishek Anand\\
       \affaddr{L3S Research Center, Leibniz University of Hannover}\\
       \affaddr{Appelstr. 9a}\\
       \affaddr{30167 Hannover, Germany.}\\
       \affaddr{\{fetahu,aanand,anand\}@L3S.de}
}

\maketitle

\begin{abstract}
Wikipedia, rich in entities and events, is an invaluable resource for various knowledge harvesting, extraction and mining tasks. Numerous resources like DBpedia, YAGO and other knowledge bases are based on extracting entity and event based knowledge from it. Online news, on the other hand, is an authoritative and rich source for emerging entities, events and facts relating to existing entities. In this work, we study the creation of entities in Wikipedia with respect to news by studying how entity and event based information flows from news to Wikipedia.

We analyze the lag of Wikipedia (based on the revision history of the English Wikipedia) with 20 years of \emph{The New York Times} dataset (NYT). We model and analyze the lag of entities and events, namely their first appearance in Wikipedia and in NYT, respectively. In our extensive experimental analysis, we find that almost 20\% of the external references in entity pages are news articles encoding the importance of news to Wikipedia. Second, we observe that the entity-based lag follows a normal distribution with a high standard deviation, whereas the lag for news-based events is typically very low. Finally, we find that events are responsible for creation of emergent entities with as many as 12\% of the entities mentioned in the event page are created after the creation of the event page.

\end{abstract}
\category{H1.1}{Information Systems}{Systems and Information Theory}[News and Wikipedia Dynamics]

\newpage

\section{Introduction}
\label{sec:introduction}

Wikipedia is the largest source of open and collaboratively curated knowledge source in the world. Introduced in 2001, it has evolved to be a very useful repository for entities, events, concepts etc. Entities and event pages are often created and collaboratively edited creating a knowledge source which is both authentic and recent. As a result, this invaluable resource has found application in information extraction and knowledge base construction, e.g.YAGO~\cite{suchanek_yago:_2007} and DBpedia~\cite{bizer_dbpedia_2009}, text categorization~\cite{wang_building_2008}, entity disambiguation~\cite{Hoffart:2011:RDN:2145432.2145521}, and entity ranking~\cite{kaptein_entity_2010}.

Owing to events being increasingly documented in online media, existing entities in Wikipedia continuously evolve and new entities are added. Moreover, online news has seen a lot of growth of late, and records important events reasonably quickly. Consequently, a high proportion of the entity pages in Wikipedia (pages devoted to named entities like persons, organizations and locations) have news articles as references, a factor which suggests that news is an authoritative source for important facts (we do a detailed analysis of the density in Section~\ref{sec:news-dyn}). 

Automatic knowledge base construction tasks can rely on news as a source or an indicator to add or update entities. First, news could be a primary source for addition of emerging entities~\cite{DBLP:conf/www/HoffartAW14}. Secondly, knowledge bases which harvest this resource need to periodically refresh their contents. They constantly deal with the natural trade-off between the cost of maintenance of a fresh and consistent state with the loss of useful information. For newsworthy entities and events, understanding this delay in appearing in Wikipedia would suitably help knowledge bases to improve their maintenance or characterize the information loss.  

In this work, we study how fast Wikipedia reacts to these real world events captured by news collections. We carry out our study on the Wikipedia revision history and the New York Times news corpus for the overlapping years between 2001 and 2007. We extract entities from the news articles and link them to the version of the entity page which is closest in time to the publication time of the article. In other words, we \emph{align} the news collection to the Wikipedia versions using entities as proxies. Next, we define \emph{lag} as the time difference between when an entity or event was reported in news and the first time it appeared in Wikipedia. This aligned resource allows us to carry out several studies which shed light on the evolution of entities and events and on how they are captured in Wikipedia. 


Specifically, we try to answer the following questions:

 \begin{itemize}
 	\item \emph{What fraction of external references in entity pages are news articles ?}

 	\item \emph{How much does Wikipedia lag behind news articles ? How has this lag evolved over time ?}

 	\item \emph{Which categories or classes of entities in news lead or lag Wikipedia ?}

 	\item \emph{How do events reported by news articles lag with the Wikipedia event pages ?}
 \end{itemize}

We perform our alignment studies on the entire English Wikipedia revision history and the New York Times collection (as the news corpora). We also consider Wikipedia's current events portal as a repository for high quality manually created resource for events. Some of the highlights of our study are:

\begin{itemize}
	\item Approximately 20\% of all external references in entity pages are news articles.

	\item Entity lag follows a distinct normal distribution and show that Wikipedia has been catching up on news ever since it was introduced. 


	\item Unlike entities, events are quickly reflected in Wikipedia as soon as they are reported in news.

	\item Events are responsible for creation of emergent entities, with 12\% of the entities mentioned in event pages being created after the creation of the event page.
\end{itemize}

 The rest of this paper unfolds as follows. In Section~\ref{sec:alignment} we introduce the experimental setup and the generated data. Section~\ref{sec:news-dyn} provides an overview of the news dynamics in Wikipedia and serves as a motivation to our study, while in Section~\ref{sec:entity-lag} and \ref{sec:event-lag} we provide thorough analysis and results on entity and event lag. In Section~\ref{sec:related-work} we review related literature relevant to our work. Finally in Section~\ref{sec:conclusion} we conclude the findings in our analysis between Wikipedia and news corpora such as NYT.

\section{Related Work}
\label{sec:related-work}
The related work and state of the art with respect to our work can be classified into the following three parts :

\textbf{Wikipedia Studies} goes into similar directions with our analysis. Kittur et al.~\cite{DBLP:bibsonomy_KCPS07} analyses the collaborator structure of Wikipedia. They further classify the collaborators into five different classes based on the number of revisions. Furthermore, they measure the population growth of the collaborators falling into the five different classes. In their paper the authors conclude an interesting observation of the shift of how content is mostly provided by collaborators with lower number of edits, due to the increased fraction of such users in the Wikipedia community structure. This, however, does not correlate with any decline of the content provided by collaborators with high number of edits, hence, is accounted to the higher fraction of low edit users. In contrast to the work from Kittur et al., we have a different focus in our analysis, namely that of entity and event lag in Wikipedia, without any distinction of the Wikipedia community structure. In \cite{Suh:2009:SNS:1641309.1641322} the authors analyze several aspects of Wikipedia's editors. They conclude that the number of edits is decreasing. Another slightly related work~\cite{Bar-Ilan:2014:TYW:2615569.2615643} analyzes the number of research papers about Wikipedia, here too they conclude that the number has been decreasing, however, papers that use Wikpedia's data has seen an increase.

Closely related work is done by Keegan et al.~\cite{keegan_hot_2011,keegan_editors_2012,hu_measuring_2007}. Their work, similarly to ours, focuses on the dynamics of Wikipedia's coverage of real world entities. In~\cite{keegan_hot_2011}, the authors consider emerging events like the T\={o}hoku catastrophe\footnote{\url{http://en.wikipedia.org/wiki/2011_T\%C5\%8Dhoku_earthquake_and_tsunami}}. In the case of such high dynamic events, it is found out that for localized Wikipedias (e.g. Japanese), the corresponding event appears only after six minutes after the event, whereas in the English Wikipedia, it appears in less than an hour. Furthermore, they analyze the co-authorship of such articles in Wikipedia. It is concluded that within Wikipedia there are sub-communities that edit articles of the same topic. As a continuation of their work, in~\cite{keegan_editors_2012} the social network structure of Wikipedia collaborators is analyzed. The analysis is based on four main hypotheses that are based on two main set of attributes, article and editor attributes, respectively. The first hypothesis validates the fact that for breaking news articles attract more editors. The second hypothesis validates the co-authorship of articles in Wikipedia from collaborators that are categorized into three main classes: \emph{Experienced, Apprentice, Non-Expert}. Significant collaborations between the three classes of collaborators is found only on \emph{contemporary} articles (articles are divided into \emph{breaking, contemporary, historical}) between \emph{apprentice} and \emph{experienced} collaborators. The third hypothesis, analyzes the editor attributes and implies that experienced editors will edit more articles than others. The third hypothesis leads to the fourth and last hypothesis. It analyzes the fact that experienced editors are more likely to contribute to similar types of articles rather than to dissimilar. Strong correlation is found for editors belonging to the \emph{apprentice} class and for most of the article types. In contrast to our analysis the work by Keegan et al. has as a main focus modeling the network structure of editors and how this reflects on the dynamics of Wikipedia and contemporary and emergent entities and events. On the other hand, in our analysis we focus on larger real world news corpora which inherently represent emerging entities and events. In addition, we also distinguish the lag for different entity types. As a last diverging point in our work, is the analysis of how entities are co-created and its impact on the entity lag.

\textbf{Entity Interlinking} tries to detect links between entities withing a knowledge base. The work done by Nunes et al. \cite{DBLP:conf/esws/NunesDCKFN13} uses social network theory measures, such as Katz index to find links between entities. This is related to our work since we analyze the co-referencing of entities within Wikipedia, their collaborator structure and interlinking with events in the Wikipedia's event portal. Such attributes of entities are used to analyze their implications on the entity lag in Wikipedia against news corpora.

\textbf{First Story Detection} typically deals with event onset identification from a stream of text. In~\cite{osborne_bieber}, Osborne et al., analyze twitter data for first story detection. Wikipedia in this case is used through its entity/event page views to filter tweets that do not represent events. The two sources of information are considered as streams which later on are mapped, by simply checking the spikes of page views for a certain entity/event in a tweet. In our case, the focus is at modeling between two sources of information, Wikipedia and NYT corpus, rather than its usage for story detection.

\section{Collection Alignment}
\label{sec:alignment}
In this section we introduce the experimental setup. To carry out our experiments we first align the two collections, Wikipedia and NYT corpus. The resulting dataset is referred to as the \emph{news-wiki aligned collection} or simply the \emph{aligned collection}. The detailed descriptions of the datasets in our experimental setup are given below:

\begin{itemize}

\item \textbf{Wikipedia} The \emph{English Wikipedia revision history}~\cite{wiki}, whose uncompressed raw data amounts to TBytes, contains the full edit history of the English Wikipedia from January~2001 to December~2013. We consider all versions of encyclopedia articles including versions that were marked as the result of a minor edit (e.g., the correction of spelling errors etc.). 
    
\item \textbf{News}  The \emph{New York Times Annotated corpus}~\cite{nyt} comprises more than 1.8 million articles from the New York Times published between 1987 and 2007. Every article has an associated publication time and we refer to this as the time of the article. Since Wikipedia was released in the year 2001 and our NYT corpora is valid until 2007 we consider sub-collections from both corpora in the time period between the years 2001 and 2007.

\item \textbf{Taxonomy} The taxonomy from the \emph{YAGO2 Ontology}~\cite{Hoffart:2011:YEQ:1963192.1963296} which combines the clean taxonomy of WordNet with the richness of the Wikipedia category system, assigning the entities to more than 350,000 classes. We also use DBPedia(resource of type \texttt{dbpedia-owl:Event}) to collect event pages along with their creation times.
\end{itemize}

\subsection{Preliminaries and Setup}\label{subsec:setup}
Before delving into detail in the lag analysis, it is necessary to introduce the entity and event notions.

\begin{entity*}\label{def:entity}
We define an entity as something which has a canonical (i.e., uniquely identifiable) representation in Wikipedia. In other words an entity represents a real world concept, e.g. \texttt{People, Organization, Location}, which might be mentioned in multiple forms in text. We refer to the Wikpedia page dedicated to a given entity as an Entity Page.
\end{entity*}

\begin{event*}\label{def:event}
It is defined as a real-world event that has a Wikipedia article, e.g. \texttt{U.S Elections 2004}. The Wikipedia article dedicated to the event is referred to as the Event Page. 
\end{event*}

We now explain in details the experimental setup. As mentioned before entities are mentioned in text, in this case news, in multiple forms and this sometimes gives rise to the problem of ambiguity, i.e., a given mention potentially refers to more than one entity. One way to resolve such ambiguities, is resolved through the task of \emph{entity linking}. \emph{Entity linking} maps such mentions of ambiguous names onto canonical entities registered in a knowledge base like DBpedia or YAGO. For this task we use \emph{TagMe!}~\cite{DBLP:journals/software/FerraginaS12}.


To maintain high accuracy of the disambiguated entities, we filter out entities with a threshold below 0.3 (values above 0.3 represent high disambiguation scores). Additionally we manually evaluate a random sample of 1000 pairs of disambiguated entities and the corresponding text snippet in the news article. The evaluation took into account whether the disambiguated entity correctly represents the entity in the text snippet. The resulting accuracy of the TagMe! tool after filtering entities, across the different entity types was on average above 0.9. After filtering the number of disambiguated entities falls to 506,151 from 722,888, with a drop of 30\% in the number of entities.

\begin{figure}[ht!]
\centering
\includegraphics[width=\columnwidth]{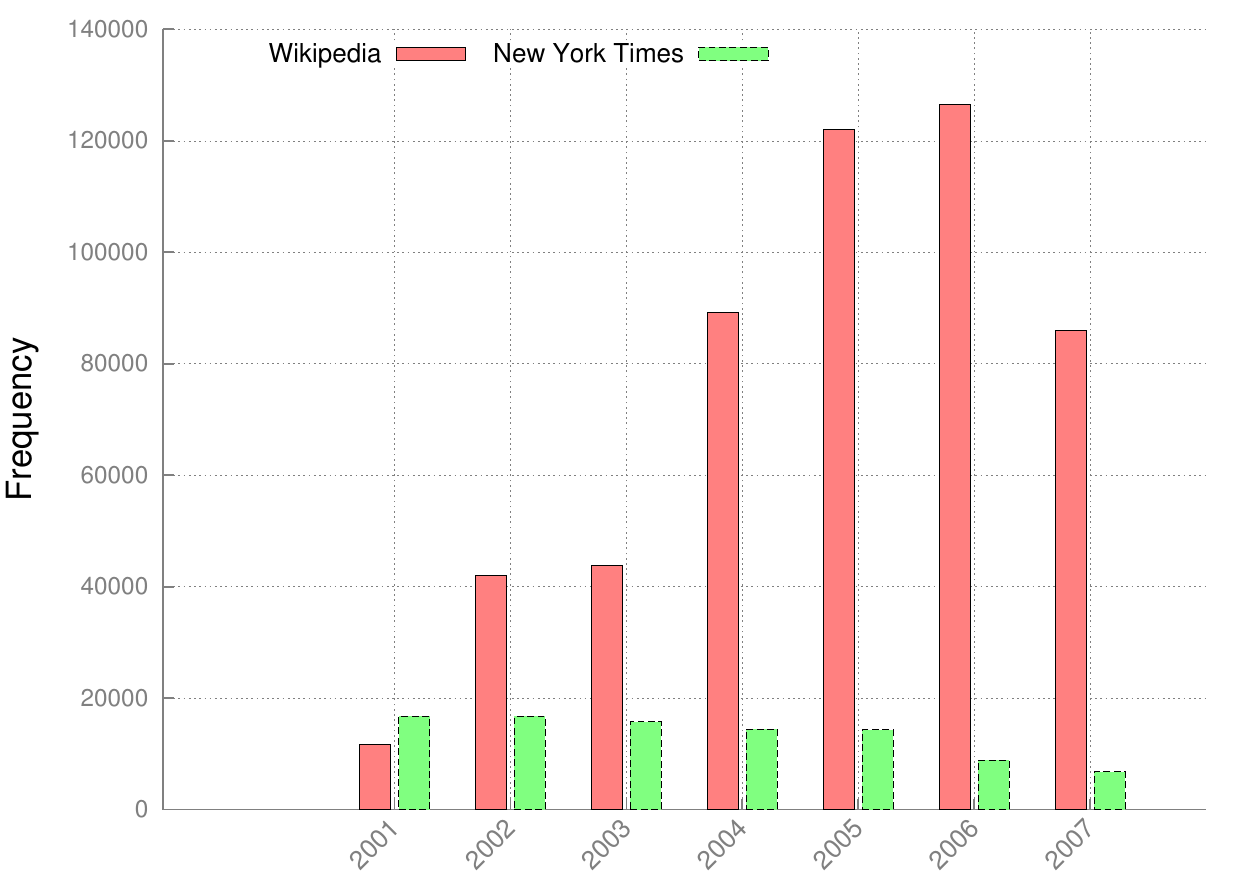}
\caption{Number of entities appearing in the corresponding years in Wikipedia, and those extracted from the named entity disambiguation process in the NYT corpora.}
\label{fig:entity_year_dist}
\end{figure}

We analyze in total 1.8 million NYT articles, resulting in approximately 506,151 distinct entities. Figure~\ref{fig:entity_year_dist} shows the distribution of extracted entities for the years 2001--2007, alongside the number of entities appearing in Wikipedia. 

The final set of entities for our experimental analysis comprises of a collection of 180,478 entities that appear only for the years 2001-2007. Furthermore, the articles are linked to the corresponding state of an entity for the specific year it appears in a NYT articles. For this purpose we make use of the JWPL~\cite{TUD-CS-2008-4}, where given an entity name and a time reference, entity revisions can be retrieved.

\begin{figure*}[ht!]
\centering
 \includegraphics[width=0.85\textwidth]{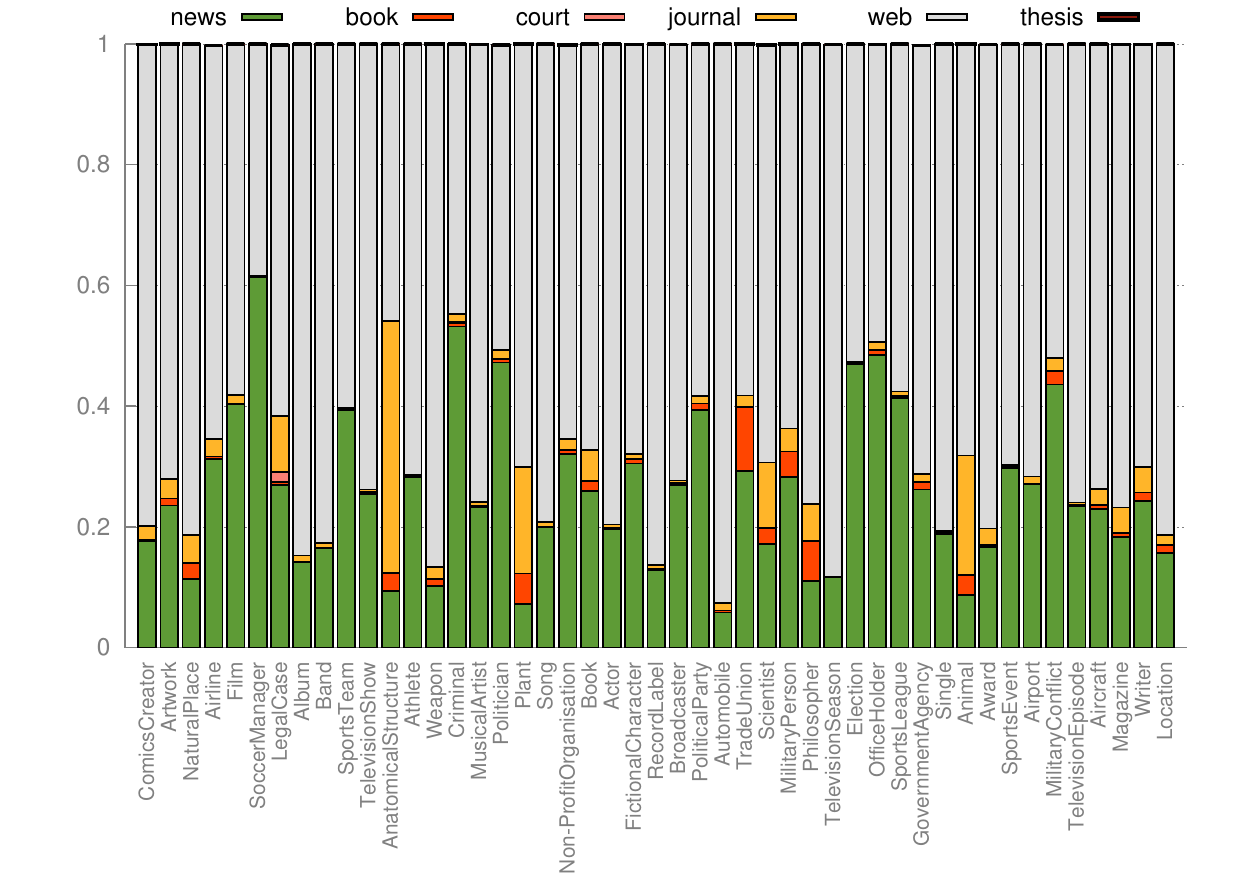}
 \caption{News Reference Density for the different entity categories. The reference density of a given reference type is measured as the fraction of references of that type over all references for the entity page.}
 \label{fig:cite_density}
\end{figure*}

\section{News Reference Density in Wikipedia}
\label{sec:news-dyn}
To start off we want to investigate how news impacts Wikipedia by studying such news references in entity pages. An \emph{entity page} refers to a Wikipedia article dedicated to an entity. Since knowledge bases are reasonable repositories of entities, we compile our set of entities from DBpedia\footnote{\url{http://wiki.dbpedia.org/Ontology}}. Entity pages, typically contain references to qualify the stated facts therein. These references are broadly classified into the following sources -- \texttt{Web, News, Book, Report and Journal, etc.} by Wikipedia\footnote{\url{http://en.wikipedia.org/wiki/Wikipedia:Citation_templates}}. We first study the distribution of news references(of type \texttt{News}) in entity pages across multiple \emph{entity categories} and the corresponding entity sections. For this experiment, we first crawled all  news articles referenced in the entity pages that are still online. This resulted in a dataset of $129,438$ available news articles out of $411,673$ news references.

\begin{news_density*}
We define the News Reference Density (NRD) of an entity page, as the fraction of news references over all references of all types in the page. Similarly reference densities of other citation types are defined.
\end{news_density*}

\begin{figure*}[ht!]
 \includegraphics[width=1.0\textwidth]{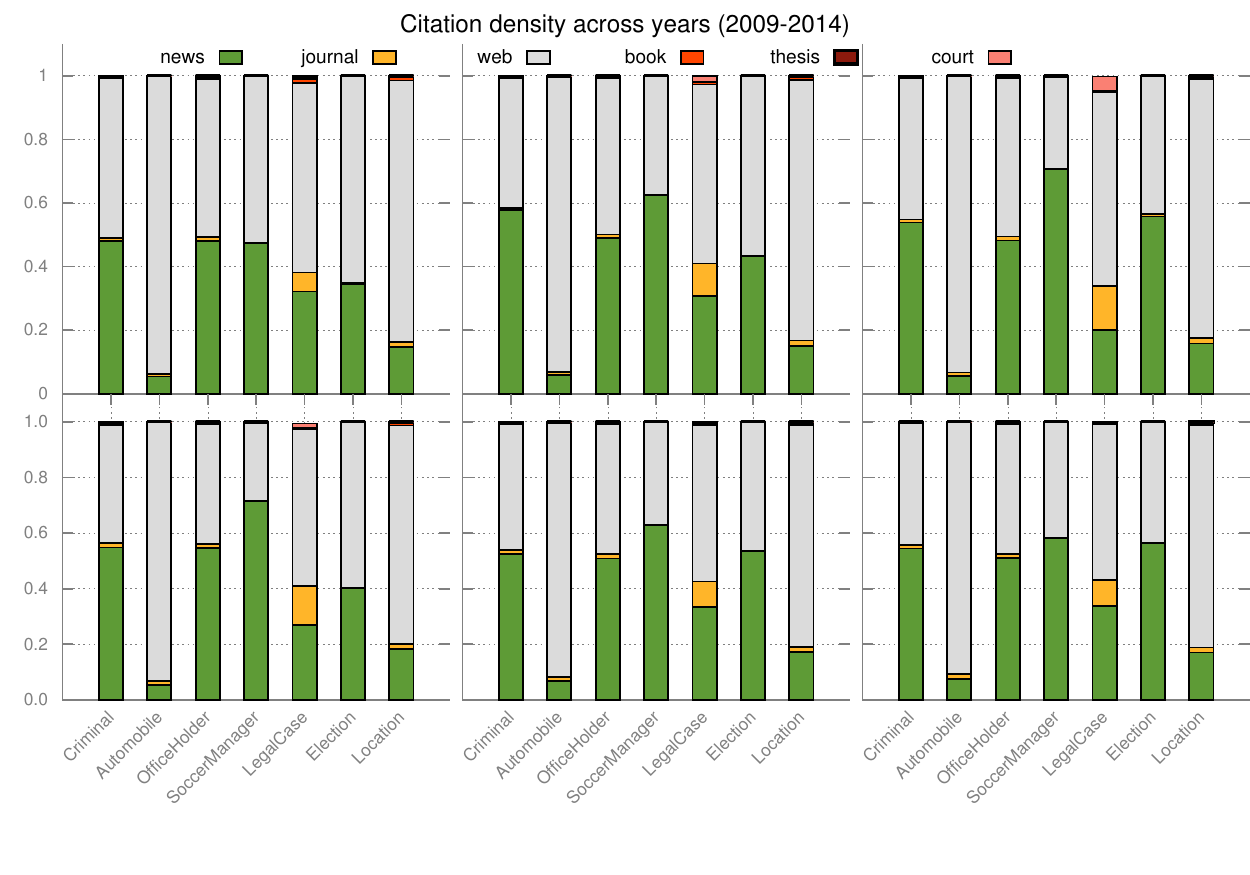}
 \caption{Reference density for the different entity categories. The plots show the reference density for years 2009-2014, in order from left to right.}
 \label{fig:domain_year_cite_density}
\end{figure*}

We observe that, as expected, most of the references are from the \texttt{Web}. However, the second most dominant type of reference are news references constituting 20\% of overall references. The NRD varies across entity categories as shown in Figure~\ref{fig:cite_density}. While the category \texttt{OfficeHolders} (mostly politicians) has a high news density, on the other hand \texttt{Bands} have high density for web references. The NRD in most cases is stable across years for the different entity categories as shown in Figure~\ref{fig:domain_year_cite_density}. However, there are slight variations on the reference density for specialized categories and the corresponding reference types, e.g. category \texttt{LegalCase} and \texttt{Court} reference types.

Taking into account the organization of Wikipedia entity pages into section, we analyze the distribution of news densities across sections in an Wikipedia entities. We observe that sections in entity pages vary considerably across categories with only some of the sections being common among categories, e.g. `\emph{Early Life}' and `\emph{Career}'. When we look at the partial contribution of the sections to the page news reference density, we observe that while `\emph{Early Life and Career}' in \texttt{Politicians} have highest NRD contribution of 64\%, the section `\emph{Sports Team}' in \texttt{Athletes} has the highest contribution of 19\%. 

\section{Entity Lag}
\label{sec:entity-lag}

\begin{figure}[ht!]
\centering
  \includegraphics[width=\columnwidth]{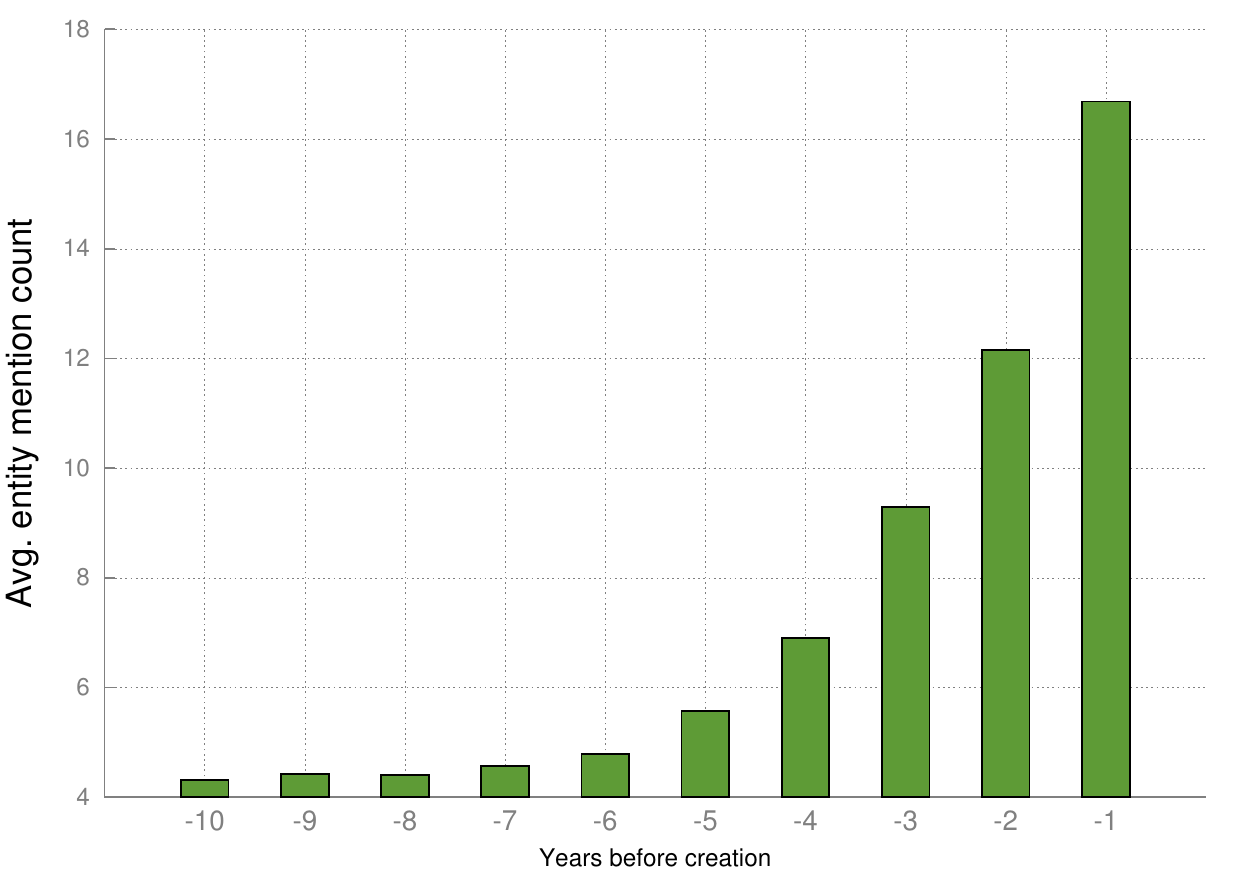}
  \caption{Entity mention counts in news articles before creation of Wikipedia entity page. Mention counts of entities peak a year before it is created in Wikipedia.}
  \label{fig:cumm_dist}
\end{figure}

\begin{figure*}[ht!]
\centering
  \includegraphics[width=1.0\textwidth]{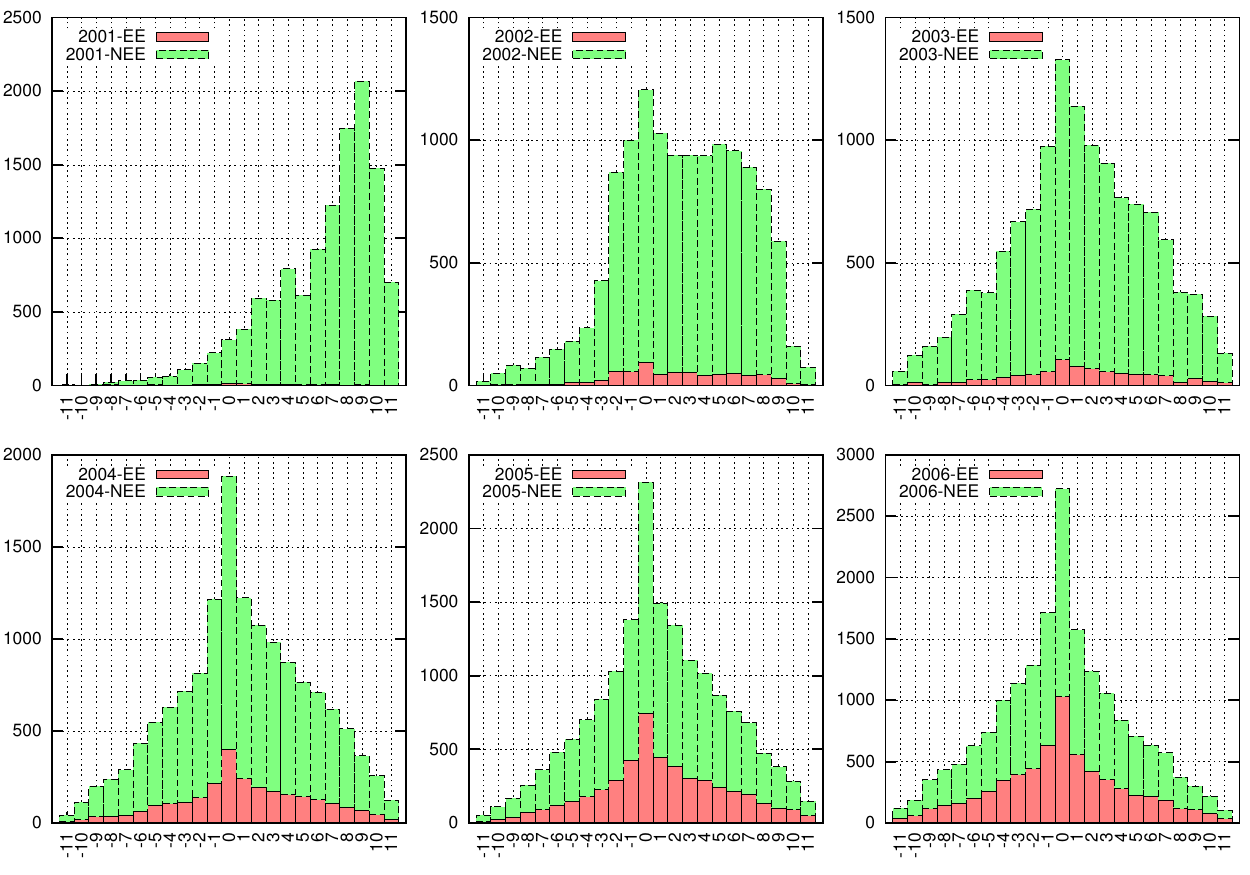}
  \caption{Entity lag in months. The emergent entities are shown in red, they are determined by filtering all entities from the subset of NYT that appear in earlier years before 2001. The y-axis is normalized using the \emph{sum} of entities having medium lag for the emerging and non-emerging entities, respectively. }
  \label{fig:hist-med-lag-dist-months}
\end{figure*}

For the concept entity we refer to the definition in Section~\ref{sec:alignment}. An entity can have multiple ways in which it can be mentioned in text. The task of resolving these mentions to the actual entities is a field of \emph{entity disambiguation}, \emph{record linkage} and \emph{entity linking}. We utilize the output of such a linking task to identify entities in our target news corpus and link them to their corresponding entity pages (see Section~\ref{subsec:setup}).

However, many of the entity pages were created at different points in time. This can be attributed to two factors: \emph{inherent popularity of the entity}, and \emph{evolution of authorship} of entity pages in Wikipedia. One explanation is that entities appearing in authoritative news sources like NYT reflect their popularity. Figure~\ref{fig:cumm_dist} shows the average entity mention distribution (in NYT) across years before the first appearance of an entity in Wikipedia. This follows the assumption that an increase of entity mentions in news sources will eventually result in the creation of an entity in Wikipedia. From Figure~\ref{fig:cumm_dist} it is obvious that shortly before the entity creation in Wikipedia, the entity is mentioned most in news. The second factor, is that Wikipedia's authorship has increased with an ever growing number of editors, hence establishing itself as a independent source of information~\cite{keegan_hot_2011}, thus entities can be created from what is deemed as important by the editors in Wikipedia.

To measure the time span between the entity mention and its creation time in Wikipedia we define the \emph{entity lag} below. 

\begin{entity_lag*} 
We define this delay of the first appearance of an entity page relative to the first appearance of the entity mention in a news article as \emph{entity lag} or simply \emph{lag} $lag(e_i)$.
$\mathbf{lag(e_i) = t_w(e_i) - t_n(e_i)}$, where $t_w(e_i)$ is the time when the first version of entity page of $e_i$ was authored and $t_n(e_i)$ is the publication time of the first mention of $e_i$ in news. 
\end{entity_lag*}

We now proceed to answer the first question of how the creation of entities in Wikipedia lag their mentions in news. We denote the entities which have an absolute lag of less than a month as \emph{low lag entities}, the ones with lag less than a year as \emph{medium-lag entities} and the rest with a lag more than a year as \emph{high-lag entities}. Figure~\ref{fig:hist-med-lag-dist-months} shows the distribution of lag in months for a period of six years. 

Second, we see that in the first year of Wikipedia the average lag was high with a majority of entities in Wikipedia lagging behind news. However, quite distinctly, the lag re-distributes towards a means of zero in the course of time into a Gaussian or normal distribution. We also see that the absolute number of entities with a lag of zeros go up, and the standard deviation reduces. The lag distribution through the years shifts to a normal distribution, with most of the entities centered around the mean, which in our case is zero. Because Wikipedia only started after 2001, we also consider the entities which were \emph{emergent} in news after 2001 (denoted by the red histogram). 

\begin{emerging_entities*}
An entity is considered as an \emph{emergent entity} (EE) if its first mention in NYT is after the time when Wikipedia was released, i.e., January 2001.
\end{emerging_entities*}

We observe that the emergent entities, much like the existing entities, have the same distribution. Since news articles are rich in political news and their coverage, we observe that emergent political topics and entities show low lag. An example is \texttt{Freedom Fries} which came into prominence in 2003 as a political euphemism for the actual French fries. On the other hand works of fiction like \texttt{The lost City} typically exhibit high lag. Similar to the non-emergent entities the lag distribution for \emph{emergent entities} is normal. In Table~\ref{tbl:sig_test} we test for normality the lag distributions for the non-emergent (NEE) and emergent entities (EE). We use the Shapiro-Wilk test for a \emph{p-value} $< 0.05$, in which case the hypothesis that the distribution is normal is rejected, otherwise for greater \emph{p-values} the hypothesis is accepted.



Based on the computed distributions, we can already provide a rough estimate of the fraction of \emph{`newsworthy'} entities, which could be missed given a maintenance period. Services that periodically update their entity repositories would lose around half of the entities if their update periods is greater than a month than if they update daily. However, there is not much gain in improving this maintenance period from 10 months to 9 months.

\begin{figure*}[t]
    \centering  
    
      \subfigure[Overall] {\label{fig:stacked_all}\includegraphics[width=0.3\textwidth]{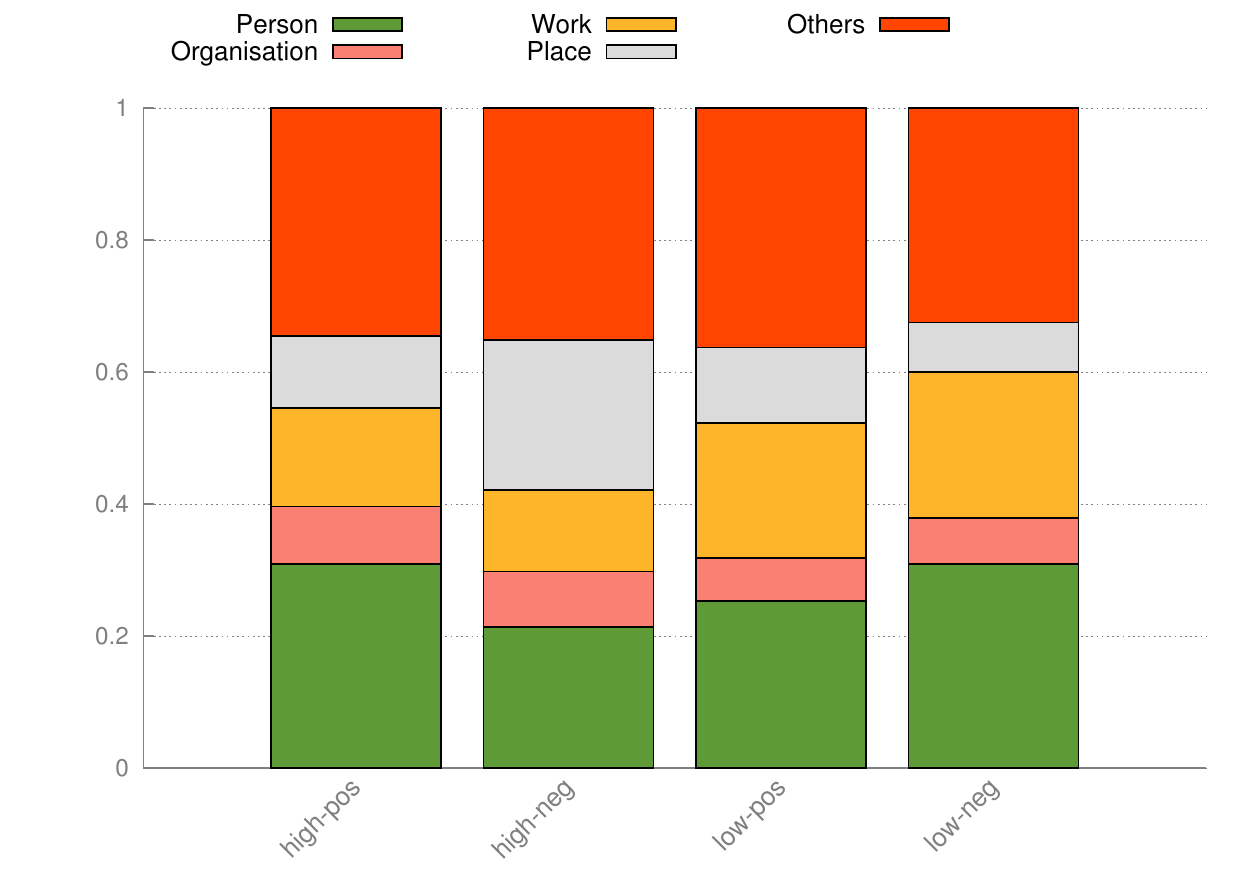}}
\hspace{0.25cm}
      \subfigure[Person] {\label{fig:stacked_person}\includegraphics[width=0.3\textwidth]{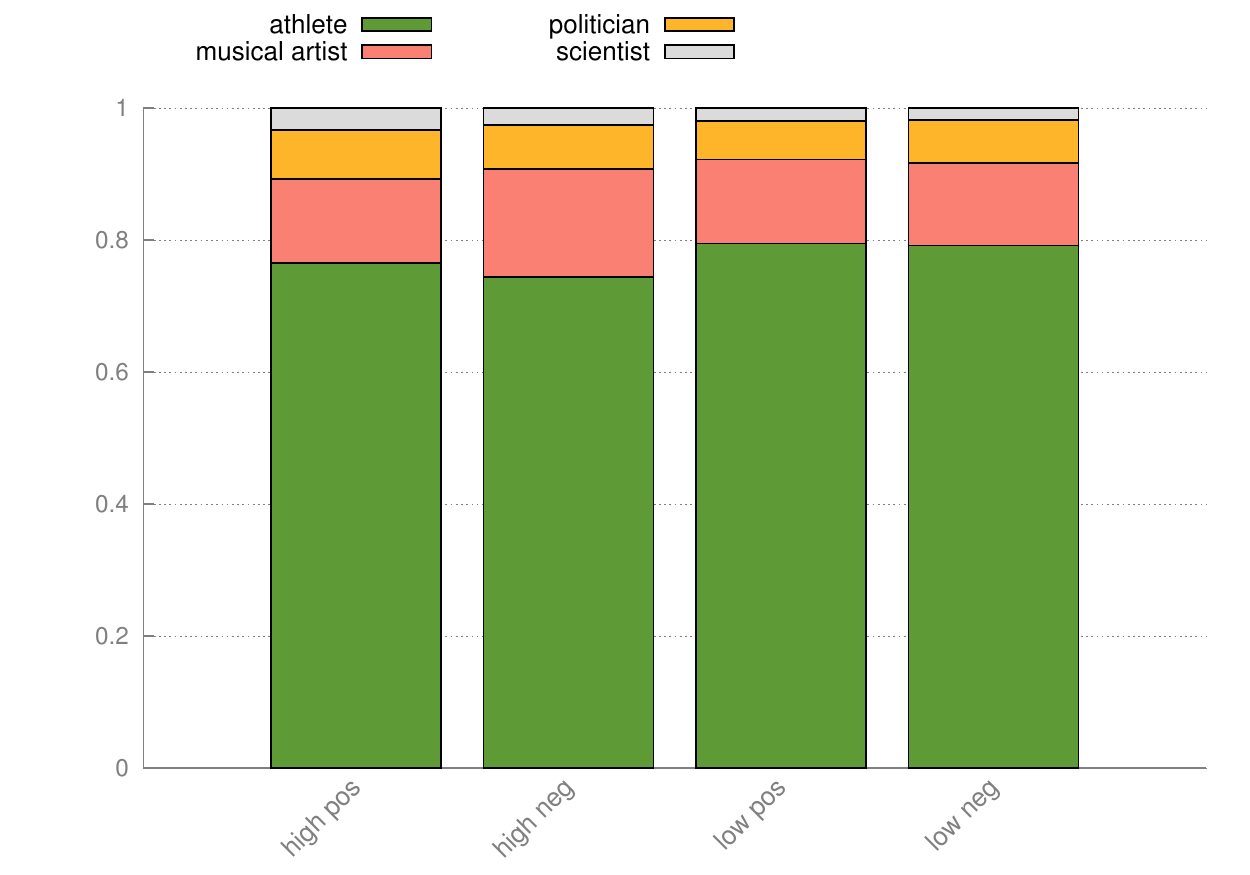}}
\hspace{0.25cm}
      \subfigure[Organization] {\label{fig:stacked_organisation}\includegraphics[width=0.3\textwidth]{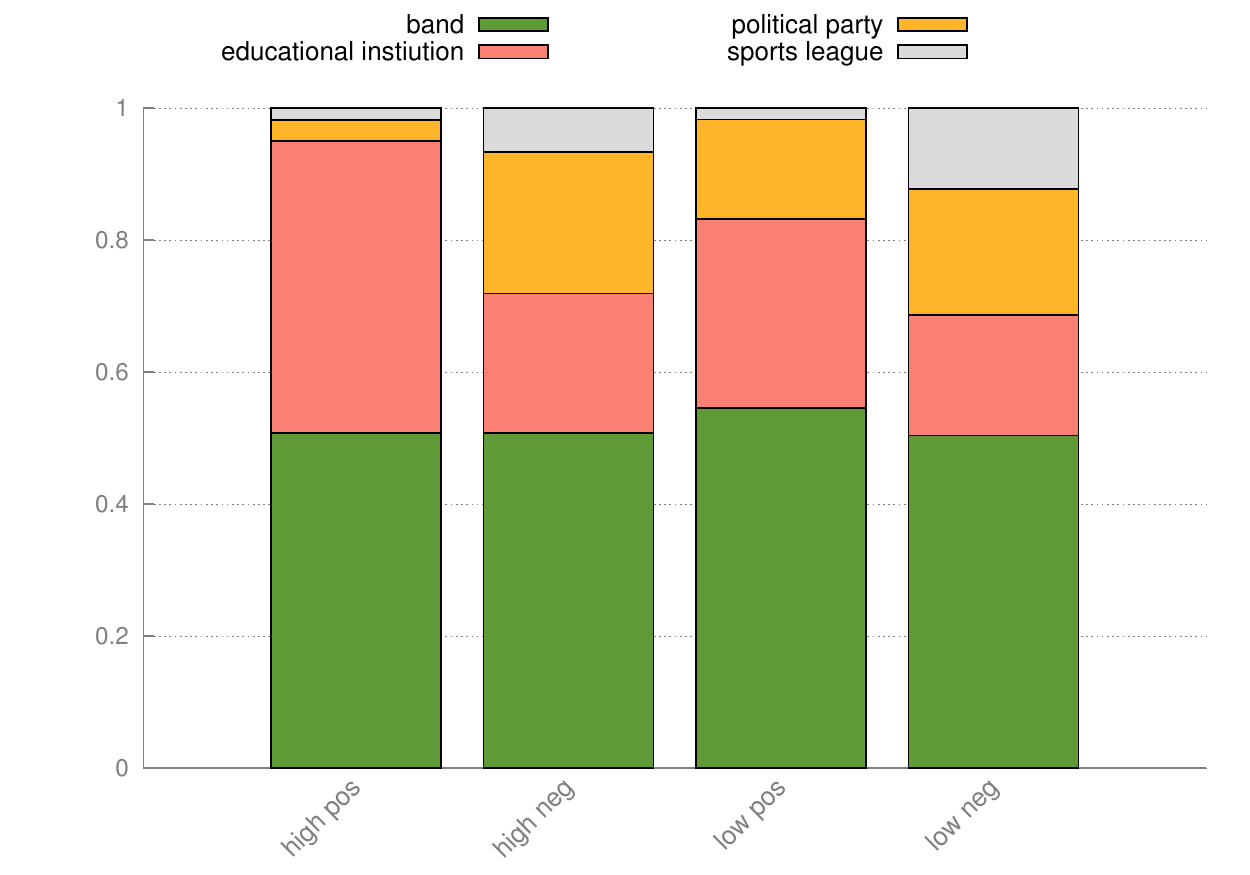}}
    \caption{Lag distribution of different entity types. The y-axis values are normalized by the sum of the overall entities falling into the different \emph{lag classes}.}
    \label{fig:entity-types}
\end{figure*}

\begin{table}[h!]
\centering
\begin{tabular}{l l l}
\toprule
\texttt{year} & \texttt{NEE} & \texttt{EE}\\
\midrule
2001 & 0.01974 &  0.00101\\
2002 & 0.01305 & 0.00155\\
2003 & \textbf{0.1177*} & \textbf{0.3585*}\\
2004 & 0.01127 & \textbf{0.2196*}\\
2005 & 0.01009 & \textbf{0.1091*}\\
2006 & 0.00269 & 0.02159\\
\bottomrule
\end{tabular}
\caption{\small{Entity lag distribution test for normality. We test whether the distributions come from a normal distribution through the \emph{Shapiro-Wilk test}. The values with * indicate that the lag distribution at the given year is normal.}}
\label{tbl:sig_test}
\end{table}

\subsection{Lag for Entity categories}

To characterize which entity classes show different lag behavior -- positive or negative , low or high -- we need to group similar entities which belong to the same semantic category. We attempt to automatically generalize sets of entities into meaningful classes based on a pre-existing taxonomy (e.g. YAGO type hierarchy).  YAGO infers class memberships from Wikipedia category names, and has integrated this information with the taxonomic backbone of WordNet  e.g, \texttt {Barack Obama} isA \texttt{US President} isA \texttt{US Politician} isA \texttt{Politician} isA \texttt{Leader} isA \texttt{Person} isA \texttt{entity}.

We first create coarse grained generalizations to obtain the major entity classes. These are \emph{Person, Work, Organization, Places, Other} and are presented in Figure~\ref{fig:stacked_all}. High-positive refers to high lag (Wikipedia lags news) whereas high-negative implies a high lead (Wikipedia leads news). It is natural to see that locations (under Places) have the highest negative lag since entity pages for many geographic locations were introduced during the early days of Wikipedia which we refer to as its \emph{bootstrapping period}.

What is interesting is that Wikipedia has a high positive lag for persons (almost 37\%) in comparison to other categories. This means that most of the emergent entities are people rather than other entity types. Looking closer into the four major subcategories of people in Figure~\ref{fig:stacked_person}, we observe that musicians tend to be mentioned in Wikipedia earlier than news and we confirm that most of them, like the locations, were also created during the bootstrapping period. We then look into the top categories of organizations in Figure~\ref{fig:stacked_organisation} and make two observations. First, all educational institutions have a high lag and secondly political parties either have a high lead or a small lag. This suggests that political parties are quite popular entities in Wikipedia while educational institutes are not. 

The entity class \emph{Work} encompasses all types of books, musical composition and movies. In general \emph{Work} is reported under low lag (around 21\%-22\%) as compared to its higher lag instances which is around around 12\%-14\%. In sum, artistic works and locations get reflected in Wikipedia sooner than other categories while Wikipedia lags news for emerging personalities. The overall distribution of entity lag is distributed as shown in Table~\ref{tbl:overal_lag}.

\begin{table}[ht!]
\centering
\begin{tabular}{l l l}
\toprule
\texttt{lag type} & \texttt{negative (lag)} & \texttt{positive (lead)}\\
\midrule
high & 57.1\% & 8\%\\
medium & 22.2\% & 11\%\\
low & 0.2\% & 1.1\%\\
\bottomrule
\end{tabular}
\caption{Absolute entity lag distributions for all lag types. The numbers are aggregated over the years 2001-2006.}
\label{tbl:overal_lag}
\end{table}

\section{Event Lag}
\label{sec:event-lag}

We now turn to studying lag in real world events as documented by news articles. Similar to the definition of entities, we define events as those which have a canonical representation via an \emph{event page} in Wikipedia. Also similar to entities, there might be more than one articles which refer to the same event. We define lag as the publication time difference between the first news article which reports the event and the Wikipedia event page. Events reported in the news can be as a reaction to an event in the past, or a build up to an upcoming event. We do not make a difference in both these cases and treat the first news article reporting the event as the inception of the event.

\begin{figure}[h!]
\centering
\includegraphics[width=0.8\columnwidth]{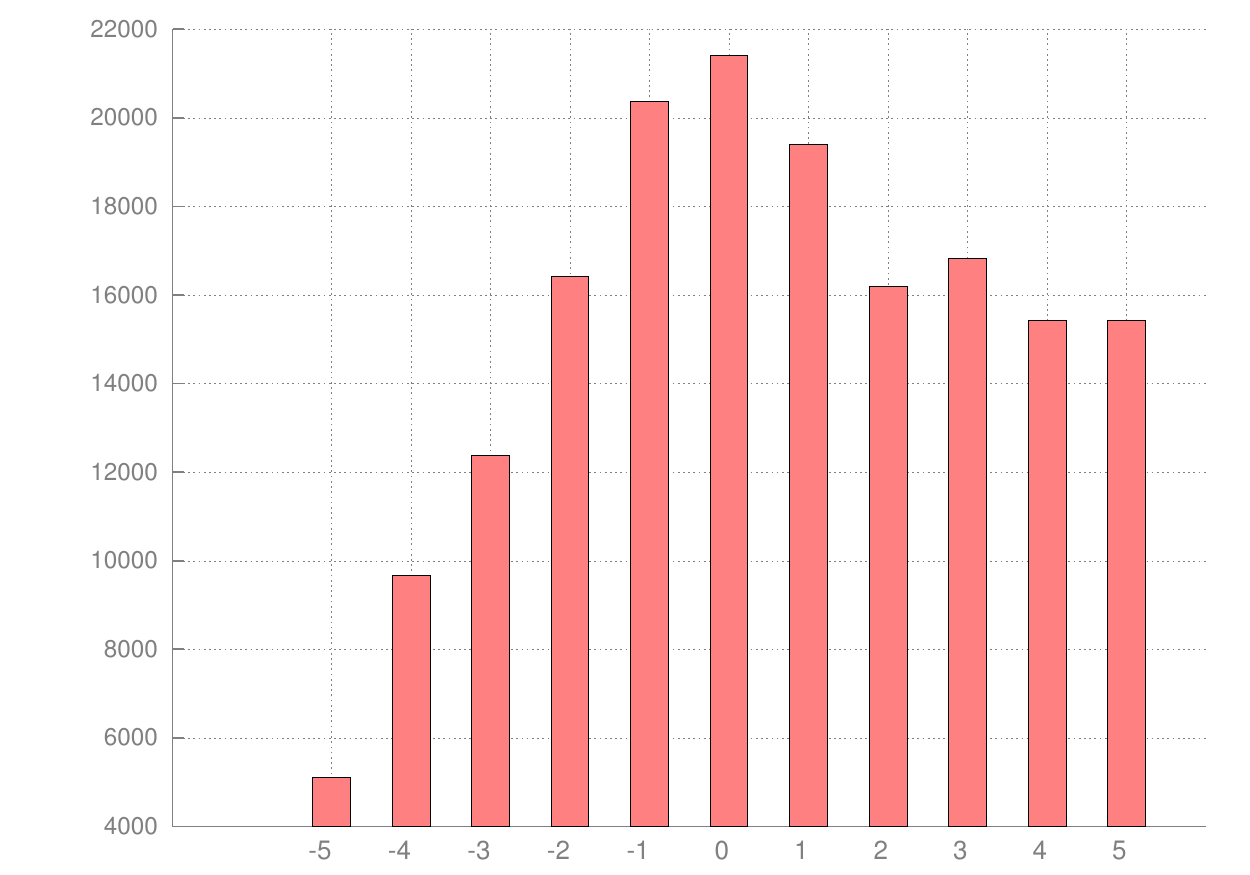}
\caption{Event news reference lag (in years) in Wikipedia. Most of Wikipedia events fall into \emph{low-lag} class, showing high dynamics of reporting real news events in Wikipedia.}
\label{fig:event_lag}
\end{figure}

\begin{figure*}[t]
    \centering  
    
      \subfigure[Emerging Entity Density] {\label{fig:EED-yearly}\includegraphics[width=0.4\textwidth]{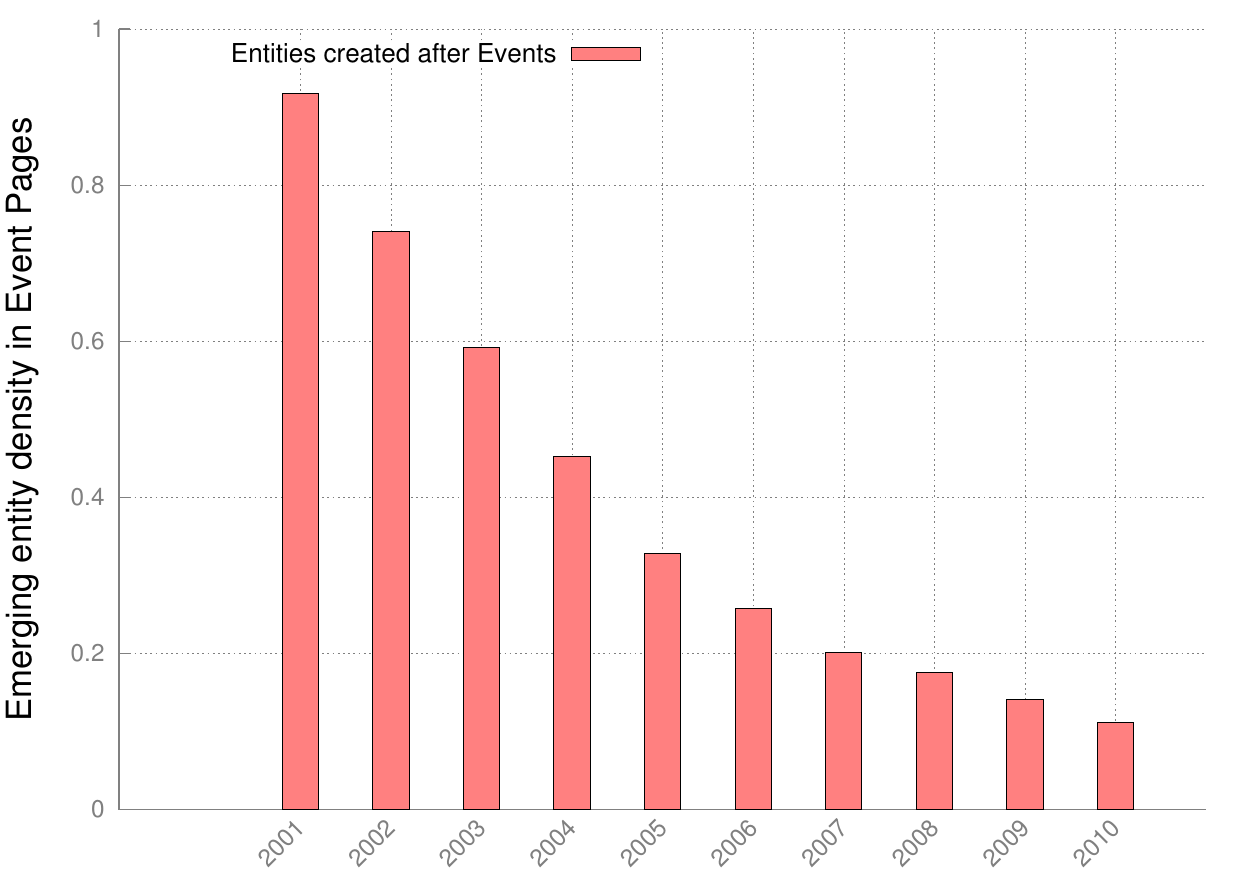}}
\hspace{0.25cm}
      \subfigure[Emerging entity categories] {\label{fig:stacked_person}\includegraphics[width=0.4\textwidth]{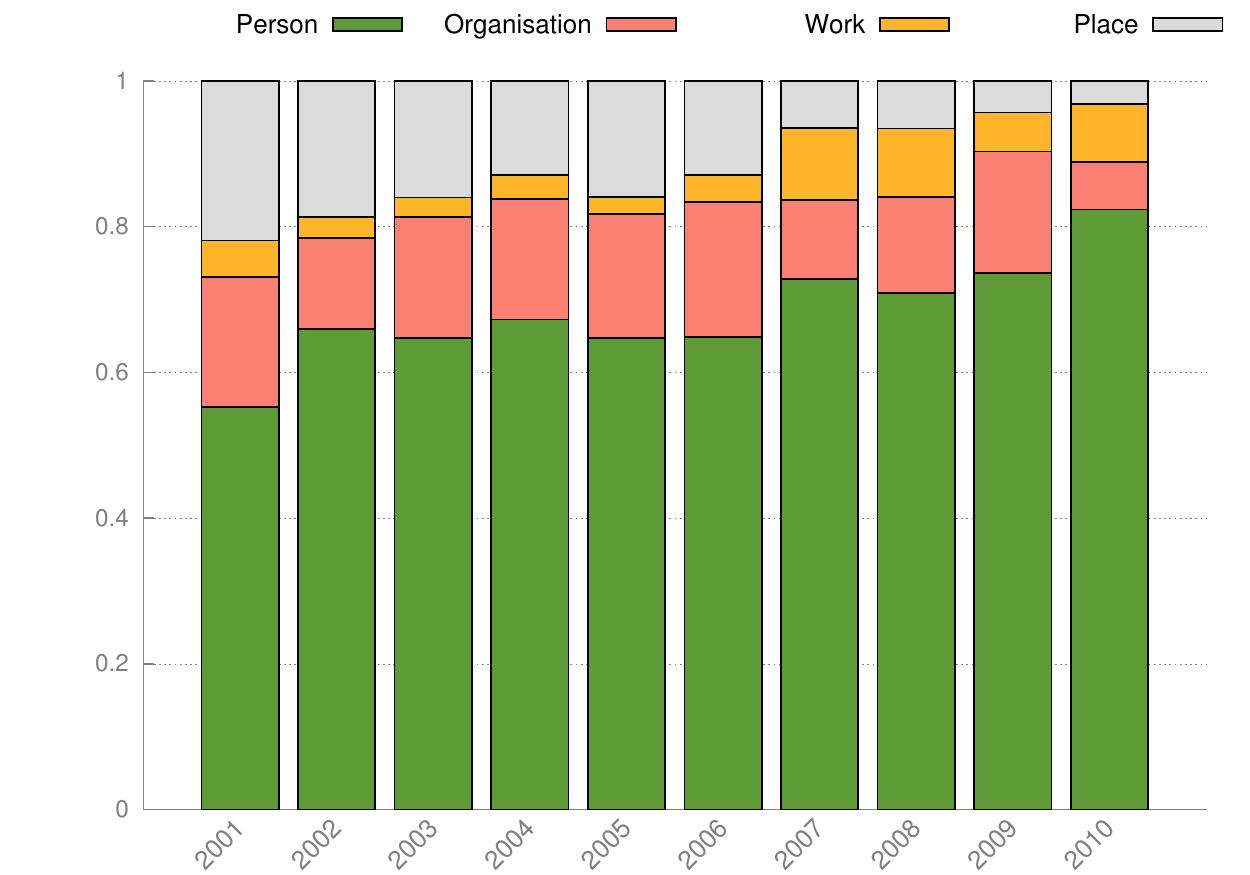}}
      
    \caption{Emerging entity density in Wikipedia event pages.}
    \label{fig:event-entity-density}

\end{figure*}

\subsection{Emerging Entities in Event Pages}
\label{sec:emerging-entities-events}

In our final experiment we study how events influence the creation of entity pages in Wikipedia. For this experiment we considered all event pages in DBpedia with their publication time (resource of type \texttt{dbpedia-owl:Event}). Unlike the previous experiments we do not rely on the NYT corpora and hence can consider the entire Wikipedia revision history.

The notion of the publication is synonymous to our earlier notion, i.e., the first time the event page was introduced in Wikipedia. We then extracted all the entities in the event page which are explicitly linked (i.e. linked to a valid Wikipedia entity page) in the most current version of the event page. Next, we compared the publication times of the entities mentioned in the events page and the event publication time. To this effect, we make a simplistic assumption about the entities mentioned in the event page: \emph{entities created after the event page are created because of this event}.

\begin{ee_ep*}
Based on this assumption, we define \emph{emerging entity density} of an event page as the fraction of entities which were created after the event page. We refer to such entities as emerging entities (note that this is different from the emergent entities present in the previous section).
\end{ee_ep*}

As an example, consider the event page of the ``\emph{Charlie Hebdo Shootings}''\footnote{\url{http://en.wikipedia.org/wiki/Charlie_Hebdo_shooting}} which was created in 7th January, 2015. The entity ``\emph{Corinne Rey}'' or ``\emph{Coco}'' \footnote{\url{http://en.wikipedia.org/wiki/Coco_(cartoonist)}} who is mentioned in the event page became popular after the event and subsequently had an entity page created five days later on 12th January.

The emerging entity density (EED) evolution from 2001-2010 is presented in Figure~\ref{fig:EED-yearly} where the y-axis represents the average emerging entity density of event pages in a given year. We have a total of 14,604 events with 179,981 entities with the exception of events from the last few years owing to the lack of event data in DBPedia for this period. We see that in the early years the EED of event pages was very high, sometimes above 80\%, meaning most of the entities mentioned in the event pages were emerging. Understandably, this declines every year resembling the phenomena of diminishing returns. However, we still see a high percentage of emerging entities in the recent event pages which point to the fact that event pages are great repositories of upcoming and emerging entities missing in the knowledge bases. We also observe that the curve, although decreasing, tends to stabilize in the recent years around 13\%. Finally, we look at the categories of emerging entities in Figure~\ref{fig:stacked_person} to find that people comprise the majority of the emergent entities consistently over the years. On the other hand, organizations were emergent between 2001-2005 but their EED contribution to event pages has been decreasing from 2006 onwards.

\section{Discussion and Conclusion}
\label{sec:conclusion}

Wikipedia is an invaluable resource documenting entities and events and is used as a important input source for constructing knowledge bases. News articles on the other hand, we find are routinely cited in Wikipedia, suggesting that they are high-quality and authoritative sources of facts about entities and events. In this work, we attempt to understand how newsworthy entities and events flow into Wikipedia by defining lag as the inter-appearance time in news and Wikipedia. We use seven years of overlapping news and the Wikipedia revision history to analyze how lag is distributed and how it has evolved over time. We see that the lag distribution is interestingly a normal distribution. 

The implications of this study is manifold. First, it shows the promise of news collections as a resource for mining emerging entities. The normal distribution of the entity lag shows that almost 50\% of the entities before occurring in Wikipedia are already mentioned in news. Hoffart et. al in~\cite{DBLP:conf/www/HoffartAW14} have initial attempts for discovering emergent entities in News and Web streams. Secondly, news is an invaluable resource for mining facts about entities and relations between entities. Our experiments on News reference density show that a high proportion of the facts and relations about entities are qualified with a news reference. Additionally our category- and section-wise analysis shows what kind of aspects of which entity-type can be found in news.

Secondly, entity and event repositories relying on Wikipedia can now quantify the degree of loss or re-calibrate their update frequencies based on the lag distribution we provide. Additionally, they can also optimize emergent entity coverage of entities by focusing on event pages. In the earlier years, Wikipedia used to lag more than news in terms of entities, while this slowly converges to a normal distribution over the years. We also observe that the lag for events is far lower than entity pages, which means they get reported far quickly. 

Thirdly, we also discover that event pages are nice containers for emergent entities with around 12\% of the entities from event pages being emergent pages. There have been studies~\cite{Suh:2009:SNS:1641309.1641322} which have reported the low growth rate in Wikipedia. We attest their finding by showing that event pages in Wikipedia, which contributed to a high number of entities, have a low emerging-entity density in the recent years. However, this low density might be because Wikipedia has eventually achieved a steady state and yields diminishing returns for new entities.

One of the limitations of the study is that new only consider the New York Times collection which might be biased towards in news coverage and hence in the entity coverage. We hope that, given the size and international nature of NYT, the results might still be representative of the overall effect of news over Wikipedia.

\section*{Acknowledgment}
This work was funded by the ERC Advanced Grant ALEXANDRIA under the grant number 339233.

\end{document}